\pdfoutput=1

\documentclass[11pt,twoside,a4paper,cmspaper,final,collab]{cms-tdr}
\def\svnVersion{239741}\def\cmsCernNo{2013-216}\def\cmsCernDate{\today}\def\cmsMessage{Published in Physical Review Letters as \href{http://dx.doi.org/10.1103/PhysRevLett.112.171801}{\doi{10.1103/PhysRevLett.112.171801.}}}

\begin{document}\cmsNoteHeader{B2G-12-012}

\hyphenation{had-ron-i-za-tion}
\hyphenation{cal-or-i-me-ter}
\hyphenation{de-vices}

\RCS$Revision: 235677 $
\RCS$HeadURL: svn+ssh://svn.cern.ch/reps/tdr2/papers/B2G-12-012/trunk/B2G-12-012.tex $
\RCS$Id: B2G-12-012.tex 235677 2014-04-07 21:00:05Z avetisya $
\newcommand{\Tft}{\ensuremath{\cmsSymbolFace{T}_{5/3}}\xspace} 
\newlength\cmsFigWidth
\ifthenelse{\boolean{cms@external}}{\newcommand{\NA}{\ensuremath{\cdots}}}{\newcommand{\NA}{\text{---}}}
\ifthenelse{\boolean{cms@external}}{\setlength\cmsFigWidth{0.85\columnwidth}}{\setlength\cmsFigWidth{0.4\textwidth}}
\ifthenelse{\boolean{cms@external}}{\providecommand{\cmsLeft}{top}}{\providecommand{\cmsLeft}{left}}
\ifthenelse{\boolean{cms@external}}{\providecommand{\cmsRight}{bottom}}{\providecommand{\cmsRight}{right}}
\ifthenelse{\boolean{cms@external}}
{\providecommand{\suppMaterial}
{Supplemental Material  [URL will be inserted by publisher]\xspace}}
{\providecommand{\suppMaterial}
{Appendix~\ref{app:suppMat}\xspace}}
\cmsNoteHeader{B2G-12-012} 
\newcommand{\usedLumi}{19.5\fbinv\xspace}
\title{Search for top-quark partners with charge 5/3 in the same-sign dilepton final state}

\date{\today}

\abstract{
A search for the production of heavy partners of the top quark with charge 5/3 is performed in events with a pair of same-sign leptons. The data sample corresponds to an
integrated luminosity of 19.5\fbinv and was collected at $\sqrt{s} = 8$\TeV by the CMS experiment. No significant excess is observed in the data above the expected
background and the existence of top-quark partners with masses below 800\GeV is excluded at a 95\% confidence level, assuming they decay exclusively to $\cPqt\PW$. This is the
first limit on these particles from the LHC, and it is significantly more restrictive than previous limits.
}

\hypersetup{%
pdfauthor={CMS Collaboration},%
pdftitle={Search for top-quark partners with charge 5/3 in the same-sign dilepton final state},%
pdfsubject={CMS},%
pdfkeywords={CMS, physics, exotic particles decaying to top quarks}}

\maketitle 
Various extensions of the standard model (SM) address the hierarchy problem, caused by the quadratic divergences in the quantum-loop corrections to the Higgs boson mass, by proposing new heavy particles. Since the largest correction arises from the top-quark loop, a  class of these models,  based on composite Higgs scenarios~\cite{DeSimone:2012fs, PhysRevD.81.075006, Dissertori:2010ug, Contino:2008hi}, predicts the existence of heavy partners of the top quark  to explain the cancellation of this correction. These ``top-quark partners'' are expected to have masses close to the
electroweak symmetry breaking scale and thus would be accessible at the CERN Large Hadron Collider (LHC), located near Geneva, Switzerland. They may also have exotic charge (5$e$/3 where $-e$ is the charge of the electron) and in this case would not contribute to the coupling of the Higgs boson to gluons~\cite{Azatov:2011qy}. Searches for such top-quark partners explore parameter space that is not excluded by  the recent observation of a Higgs boson with  properties consistent with those of the SM Higgs particle~\cite{Aad:2012tfa, higgspaper, Chatrchyan:2013lba, Aad:2013wqa, Aad:2013xqa, Chatrchyan:2012jja, Chatrchyan:2013iaa, Chatrchyan:2013mxa, Chatrchyan:2014vua}. Theoretical predictions suggest that searches in the mass
region from 500\GeV to 1.5\TeV present the greatest potential for discovery at the LHC~\cite{Contino:2006qr,PhysRevD.81.075006}.

This Letter presents a search for exotic top-quark partners using LHC $\Pp\Pp$ collision data collected by the Compact Muon Solenoid (CMS) experiment at a center-of-mass energy
$\sqrt{s}=8$\TeV. The analysis is based on a data sample corresponding to an integrated luminosity of $\usedLumi$.
We look for the $\Tft$, an exotic top-quark partner with charge 5$e$/3. We assume that the
$\Tft$ is pair-produced via either gluon fusion or quark annihilation and decays via $\Tft \to\cPqt\PWp$ followed by $\cPqt
\to \PWp \cPqb$ (charge conjugate modes are implied throughout this Letter).  Single  $\Tft$ production is not considered because it is more
model dependent and presents a different event topology~\cite{PhysRevD.81.075006}.

We focus on the dilepton final state wherein, for one or both of the $\Tft$, its two $\PW$ bosons both decay into leptons, which will have the same charge. Due to
the presence of the two bottom quarks and the possibility of hadronic decays for one of the top-quark partners, this final state also includes significant jet activity.
The leptons considered in this analysis are electrons and muons. The presence of leptons with the same electric charge (same-sign leptons) distinguishes this process from $\ttbar$, making the contribution of
the latter comparable to backgrounds with much smaller cross sections: $\ttbar \PW$, $\ttbar\PW\PW$, $\ttbar\cPZ$, $\PW\PW\PW$, and same-sign $\PW\PW$. Because of its large
cross section, $\ttbar$ still contributes to the overall background through instrumental effects such as charge misidentification in dilepton decays, as well as through
$\ttbar$ events where the $\PW$ boson from one top quark decays leptonically and the second lepton arises from a b-quark decay. Additional processes that contribute to the
expected background include QCD multijets, $\PW/\cPZ$+jets, and dibosons (\PW\cPZ\ and \cPZ\cPZ).  A previous search using a signature of same-sign leptons, multiple jets,
and missing transverse energy was performed by the CDF experiment and excludes $\Tft$ masses below 365\GeV at the 95\% confidence level
(CL)~\cite{Aaltonen:2009nr}. The CDF collaboration also set a limit on the production of exotic quarks with charge $-4e$/3~\cite{Aaltonen:2010js}.

The central feature of the CMS apparatus is a superconducting
solenoid of 6\unit{m} internal diameter, providing a magnetic field of 3.8\unit{T}.
Within the superconducting solenoid volume are a silicon pixel and strip tracker, a
lead tungstate crystal electromagnetic calorimeter (ECAL), and a brass and scintillator
hadron calorimeter. Muons are measured in gas-ionization detectors embedded
in the steel flux return yoke outside the solenoid. In addition, the CMS detector has extensive forward calorimetry.
A more detailed description of the CMS detector can be found elsewhere~\cite{Chatrchyan:2008zzk}.

Simulation of the pair production of top-quark partners was performed with the \MADGRAPH~5.1.1~\cite{Alwall:2011st} event generator, and twelve samples corresponding to values
of the $\Tft$ mass from 350\GeV to 1\TeV were produced. The \PYTHIA~6.426~\cite{Sjostrand:2006za} generator was used for parton showering, hadronization,
and simulation of the underlying event. The CTEQ6L~\cite{Pumplin:2002vw}
parton distribution functions were used and the \PYTHIA  parameters for the underlying event were set to the  Z2*~\cite{UEPAS} tune. The detector response was modeled using \GEANTfour~\cite{GEANT}.
The next-to-next-to-leading-order cross section for $\Tft$ pair production was found using \textsc{Top++}~\cite{Czakon:2013goa} to vary from 5.3\unit{pb} at the mass of 350\GeV to 3.4\unit{fb} at the mass of 1\TeV. The uncertainty on the cross section in the mass range used for the analysis is about 5\%.

The background processes include $\ttbar$, $\PW(\to\ell\nu)$+jets, $\cPZ/\gamma^*(\to\ell\ell$)+jets, and dibosons ($\PW\cPZ$ and $\cPZ\cPZ$). Additional low-rate SM processes were also considered: $\PW^{\pm}\PW^{\pm}$, $\PW\PW\PW$, $\ttbar \PW$, $\ttbar\PW\PW$, and $\ttbar \cPZ$.   These were all simulated by means of \MADGRAPH with \PYTHIA used for hadronization. Next-to-leading-order or, where available, next-to-next-to-leading order cross sections were employed~\cite{Garzelli:2012bn,MCFM}.

For all the simulated samples, the additional proton-proton interactions in each beam crossing (pileup) were modeled by superimposing minimum bias interactions
(obtained using \PYTHIA with the Z2* tune) onto simulated events, with the multiplicity distribution matching the one observed in data.

This analysis relies on the reconstruction of three physical objects: electrons, muons, and jets. The events are reconstructed using the CMS ``particle-flow'' event description algorithm~\cite{pf,pf2}.
Candidate events are required to have at least two leptons with the same charge that are within the detector acceptance ($\abs{\eta} <2.4$) and to
have passed triggers based on dielectrons, dimuons, or electron-muon combinations.
Here, $\eta$ is the pseudorapidity defined as $\eta \equiv -\ln[\tan(\theta/2)]$, where $\theta$ is the polar angle with respect
to  the counterclockwise beam direction. Candidate events must also have at least one good reconstructed primary vertex matched to the track origins of the two selected leptons.
Candidate electrons are reconstructed using energy deposits in the ECAL and a track from the silicon detectors. The shape of the shower in the ECAL must be
consistent with that of an electron and the shower must be well matched to the extrapolated track. Additional requirements are imposed to reject electrons produced by photon conversions. The charge of each electron candidate is measured using three different methods.  Two of the measurements are based on two different tracking
algorithms: the standard CMS track reconstruction algorithm~\cite{ctf} and the Gaussian Sum Filter algorithm~\cite{gsf}, optimized to take into account the possible emission
of bremsstrahlung photons in the silicon tracker. The third measurement is based on the relative position of the calorimeter cluster and the projected track from the pixel
detector. All three measurements are required to be consistent with the electron hypothesis. Muon candidates are required to be reconstructed by both the silicon tracker and the muon system and the combined fit of
the track must be of good quality ($\chi^2$ per degree of freedom less than ten).

All selected leptons are required to be isolated. The isolation for each lepton is estimated by first computing the scalar sum of the transverse momenta of all neutral and
charged reconstructed particle candidates, except the lepton itself, within a cone of size $\DR \equiv \sqrt{(\Delta \phi)^{2}+(\Delta \eta)^{2}}$ around the lepton, where $\phi$ is
the azimuthal angle. This sum is then divided by the transverse momentum (\pt) of the lepton to calculate the relative isolation ($I_{R}$). The values for the cone size and
the maximum allowed $I_{R}$
are $\DR = 0.3\,(0.4)$ and $I_{R} = 0.15\,(0.20)$ respectively for electrons (muons), as determined by optimization studies for CMS top-quark related analyses.
An event-by-event correction is applied to the computation of the lepton isolation in order to account for the effect of pileup.
Scale factors to correct for imperfect detector simulation are obtained using the tag-and-probe method~\cite{CMS:2011aa} for lepton identification and isolation as a function of
lepton \pt and $\eta$.
In addition, we define a category of ``loose'' leptons with some of the isolation and identification requirements relaxed. The $I_{R}$ threshold for these leptons is
increased from 0.15 to 0.60 for electrons and from 0.20 to 0.40 for muons.

For the range of $\Tft$ masses accessible at $\sqrt{s} = 8$\TeV, the analysis exploits advanced  techniques in jet reconstruction for identifying highly boosted top quarks and $\PW$ bosons that decay hadronically.
 In particular, if the top quarks are highly boosted
($\pt > 400$\GeV), their decay products are collimated and merged into one jet. We use a ``top-quark tagging'' algorithm based on identifying jet substructure~\cite{CMS-PAS-JME-09-001} to
reconstruct such merged top-quark jets. Jets are clustered using the Cambridge--Aachen algorithm~\cite{Wobisch:1998wt,Dokshitzer:1997in}, as implemented in \textsc{FastJet} version 3~\cite{Cacciari:2011ma},
with a distance parameter of $R=0.8$ in $\eta$-$\phi$ space (CA8 jets). The CA8 top-quark jets are required to have $\pt > 400$\GeV and more than two subjets found by the top-quark tagging algorithm.
The jet mass must be consistent with the mass of the top quark and the minimum pairwise mass of the three highest $\pt$ subjets is required to be greater than 50\GeV.

The decay products of $\PW$ bosons from the $\Tft$ decay or from a highly boosted top quark, for which the $\cPqb$ quark is reconstructed independently, may also merge into a single jet.  We use a ``jet pruning'' algorithm~\cite{CMS-PAS-JME-10-013} to identify the hadronic decay of such $\PW$ bosons. This algorithm also uses CA8 jets as inputs with the pruning parameters taken from the original theoretical papers~\cite{Ellis:2009su,Ellis:2009me}. CA8 \PW-boson jets are required to have $\pt > 200$\GeV, exactly two subjets, and their mass must be consistent with that of the $\PW$ boson~\cite{CMS-PAS-JME-13-006}.

To account for $\PW$ bosons and top quarks that are not highly boosted, jets are also reconstructed using the anti-\kt algorithm~\cite{Cacciari:2008gp} with a distance parameter of 0.5 (AK5). These jets are required to have $\pt > 30$\GeV. If an AK5 jet overlaps with a top-quark jet or a \PW-boson jet ($\DR < 0.8$), the AK5 jet is discarded.

All of the above categories of jets are required to have $\abs{\eta} < 2.4$ and particle-flow jet identification~\cite{JINSTJetPaper}. Jet energy corrections are applied to
account for residual non-uniformity and non-linearity of the detector response. Jet energies are also corrected by subtracting the average contribution of particles from
pileup~\cite{Fastjet1,Fastjet2}. For the simulated samples, additional smearing is applied to the jet $\pt$ ($\sim$7--19\% depending on $\eta$) in order to reproduce the jet energy
resolution observed in data. All jets must be $\DR\ge0.3$ away from the selected leptons and, as mentioned above, $\DR\ge0.8$ away from any other jet. A correction to account for
differences in the identification efficiency of \PW-boson and top-quark jets between  data and simulation is applied~\cite{Chatrchyan:2013lca}.

The signal selection, optimized to yield the best signal sensitivity, requires:
\begin{itemize}
\item At least two isolated same-sign leptons as defined above with $\pt > 30$\GeV. Between each lepton and every top-quark jet, we require $\DR > 0.8$.
\item Dilepton \cPZ-boson veto: $M(\Pe\Pe)<76\GeV$ or $M(\Pe\Pe)>106\GeV$. This selection applies only to the dielectron channel. If the muon charge is mismeasured,
its momentum will also be mismeasured so a selected muon pair from a $\cPZ$ boson will not fall within this invariant mass range.
\item Trilepton \cPZ-boson veto: $M(\ell\ell)<76\GeV$ or $M(\ell\ell)>106\GeV$ where $M(\ell\ell)$ is the invariant mass of
either one of the selected leptons and any other same flavor opposite-sign lepton in the event with $\pt > 15$\GeV that satisfies the loose lepton
criteria.
\item $N_c \ge 7$, where $N_c$ is the number of constituents identified in the event. For the purpose of this selection, each AK5 jet and each lepton counts as one constituent.
Since a \PW-boson jet is assumed to correspond to a \PW\ boson, each such jet counts as two constituents, corresponding to the \PW-boson decay products.
Likewise, each top-quark jet represents a top quark and counts as three constituents.
\item $\HT > 900\GeV$, where \HT is the scalar sum of the \pt of all selected jets and leptons in the event.
\end{itemize}
With these criteria, the signal efficiency is 10--13\% for $\Tft$ masses between 750 and 1000\GeV.

The backgrounds associated with this analysis fall into three main categories. First, they may originate from SM processes leading to prompt, same-sign dilepton signatures,
including
diboson production ($\PW\cPZ$ and $\cPZ\cPZ$), $\ttbar\PW$, $\ttbar\PW\PW$, $\ttbar \cPZ$, $\PW^{\pm}\PW^{\pm}$, and $\PW\PW\PW$. The contribution of these backgrounds is obtained from simulation.

The second category consists of events from processes with prompt, opposite-sign leptons, such as $\ttbar$ and Drell--Yan production, in which one of the leptons is misreconstructed with the wrong charge, leading to a
same-sign dilepton final state. For muons in the \pt range typical of the dominant backgrounds, the charge misidentification rate is extremely small (of order $10^{-4}$) and its contribution to the background is
negligible~\cite{susy2011}. For electrons, the charge misidentification probability (${\sim}10^{-3}$) is derived from a data sample dominated by Drell--Yan events obtained
by selecting dileptons with an invariant mass consistent with originating from the \cPZ\ boson, using the ratio of same-sign \cPZ\ boson candidates to the total number of candidates.
The number of expected same-sign events due to charge misidentification is then estimated by considering the total number of events passing the
full selection, but having oppositely charged leptons. These events are weighted by the charge misidentification probability parametrized as a function of the electron $\pt$ and $\eta$ to obtain the contribution of this
background type.

The third category consists of events with one or more ``non-prompt leptons''. This is the primary instrumental background arising from jets being misidentified as leptons and
non-prompt leptons passing tight isolation selection criteria. This contribution is estimated using the ``Tight-Loose'' method described
in~\cite{Chatrchyan:2011wba}.
``Tight'' leptons have the same definition as those used in the analysis, whereas ``loose'' leptons are defined earlier.
The background is estimated by using events with one or more loose leptons weighted by the ratios of the numbers of tight leptons to the numbers of loose leptons
expected for prompt and non-prompt leptons. The ratio for prompt leptons is determined from Drell--Yan events where the invariant mass of the leptons is within 10\GeV of the \cPZ\ boson mass.
The non-prompt ratio is determined from a sample enriched in background by requiring exactly one lepton, low missing transverse energy ($\MET < 25$\GeV), low transverse mass
($M_\mathrm{T} < 25$\GeV), and at least one jet (the ``away-jet'') with $\pt > 40\GeV$ and $\DR > 1.0$ with respect to the lepton. The transverse mass is defined as $M_\mathrm{T} \equiv \sqrt{2 p_\mathrm{T}^\ell \MET (1 - \cos \Delta \phi) }$ where  $\Delta \phi$ is the angle between the lepton transverse momentum ($p_\mathrm{T}^\ell$) and the direction associated with
$\MET$.

The systematic uncertainties that affect the signal and background acceptance include uncertainties in the efficiency of the trigger (1\%), lepton reconstruction and identification efficiency
(1\% per lepton), pileup, and jet energy scale (JES). The uncertainties due to JES and pileup are obtained by varying the respective quantities in simulation.
For the signal, the JES and pileup uncertainties in the acceptance correspond to 2\% and 3\%, respectively. For the simulated backgrounds, they range from 3 to 6\% depending on the sample.
In addition, we assign a constant 3\% uncertainty due to the JES of CA8 jets for all simulated samples~\cite{Chatrchyan:2013lca}.
The dominant uncertainty in the expected event yields due to backgrounds derived from simulation is the overall normalization uncertainty. The $\PW\cPZ$ (17\%), $\cPZ\cPZ$ (5.1\%),
and $\ttbar\PW$ (32\%) normalization uncertainties are taken from Refs.~\cite{WZAtlas2012}, ~\cite{MCFM}, and~\cite{Campbell:2012dh}, respectively. For the other rare
backgrounds, we assume a conservative normalization uncertainty of 50\%~\cite{Chatrchyan:2012paa}. An uncertainty of 20\% is assigned to the background contribution from charge misidentification, based
on the difference in the charge misidentification rate between Drell--Yan data and $\ttbar$ simulation.
Following Ref.~\cite{susy2011}, we also assign a conservative additional systematic uncertainty of 50\% in the estimation of backgrounds due to non-prompt leptons. This uncertainty is
based on closure tests using $\ttbar$ and $\PW$+jets simulated samples and takes into account variations due to the away-jet $\pt$ and the flavor composition of the background, thus
also accounting for any potential dependence on kinematic parameters that alter the background composition (such as \HT).
We also include a 2.6\% uncertainty due to the luminosity~\cite{CMS-PAS-LUM-13-001} for all event yields that are derived from simulation.

\begin{figure}[tbp]
\begin{center}
\includegraphics[width=0.48\textwidth]{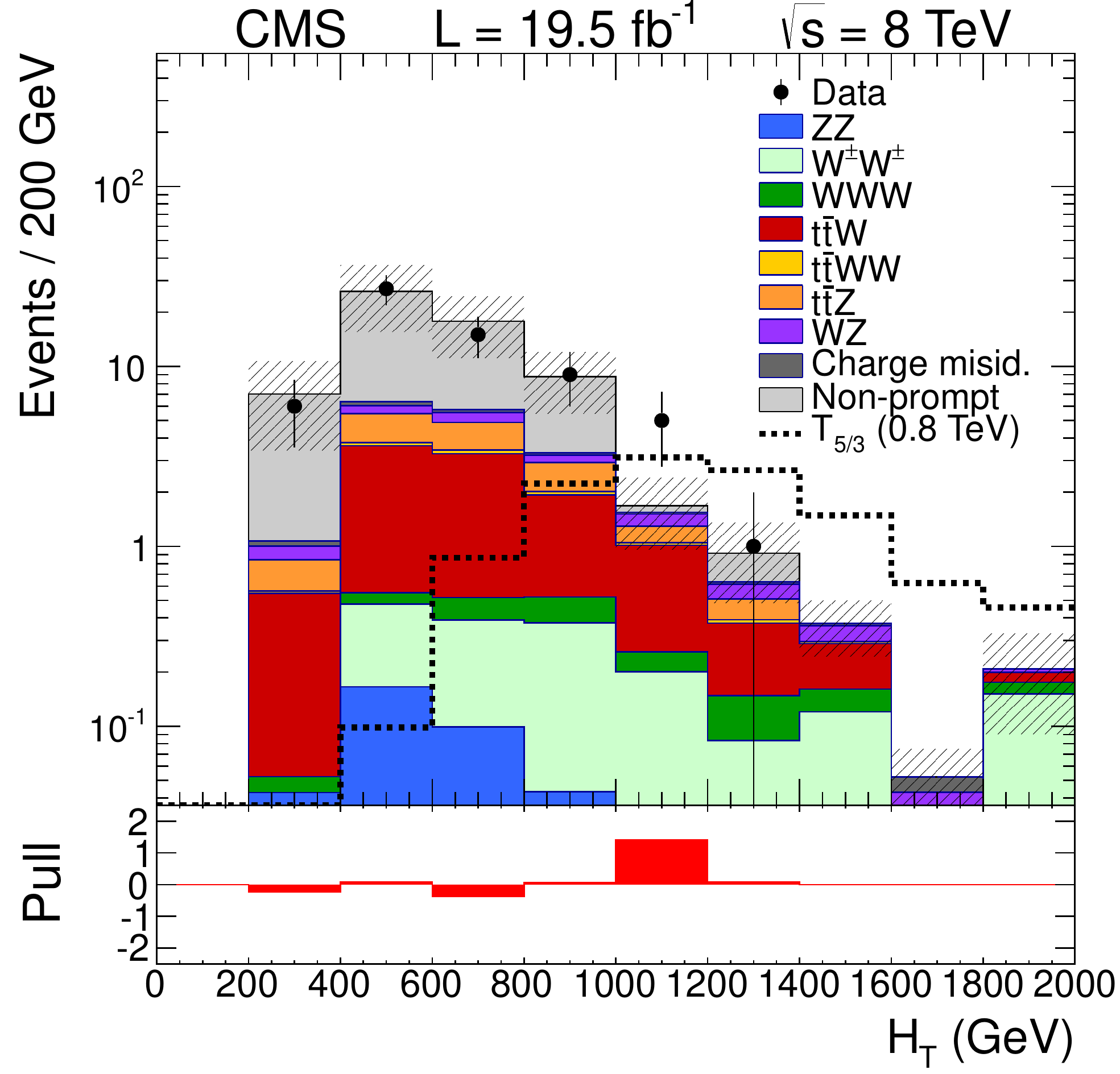}
\caption{The distribution of \HT for all channels combined after the full selection except for the \HT requirement itself.
The shaded band represents the total uncertainty in the predicted backgrounds. The final bin includes all overflow events.
The pull is defined as the difference between the observed and expected values divided by the total uncertainty.}
\label{fig:HT}
\end{center}
\end{figure}

The final numbers of observed and expected events are reported in Table~\ref{tab:FinalYield} for each of the three lepton channels ($\Pe\Pe$, $\Pe\Pgm$, and $\Pgm\Pgm$) and their combination.
Figure~\ref{fig:HT} shows the \HT distribution for all channels combined.

\begin{table}[!htbp]
\centering
\topcaption{Summary table of expected and observed numbers of events for all channels. The background is composed of the same-sign component,
         the contribution due to charge misidentification, and that due to misreconstructed leptons. All systematic uncertainties are included. Also shown is the expected contribution from a $\Tft$ with mass 800\GeV. }
\begin{scotch}{lcccc}

Channel                       & $\Pe\Pe$          & $\Pe\Pgm$          &  $\Pgm\Pgm$       &   All          \\
\hline
Same-sign                     & 0.8  $\pm$ 0.2  & 1.9  $\pm$ 0.4  & 1.3 $\pm$ 0.3 & 4.0 $\pm$ 0.8  \\
Chrg. misid.                 & 0.06 $\pm$ 0.02 & 0.04 $\pm$ 0.01 &       ---       & 0.11 $\pm$ 0.02 \\
Non-prompt                    & 1.9  $\pm$ 1.2  & 0.6  $\pm$ 0.9  & 0.3 $\pm$ 0.6 & 2.8 $\pm$ 1.9  \\
\hline
Tot. bkgnd              & 2.7  $\pm$ 1.3  & 2.5  $\pm$ 1.0  & 1.6 $\pm$ 0.7 & 6.8 $\pm$ 2.1  \\
\hline
Obs. events               &        0        &         6       &       3       &      9         \\
\hline
$\Tft$  & 2.1 $\pm$ 0.1   & 4.7 $\pm$ 0.3   & 2.8 $\pm$ 0.2 & 9.7 $\pm$ 0.5  \\
\end{scotch}
\label{tab:FinalYield}
\end{table}

\begin{figure}[htbp]
\centering
\includegraphics[width=0.48\textwidth]{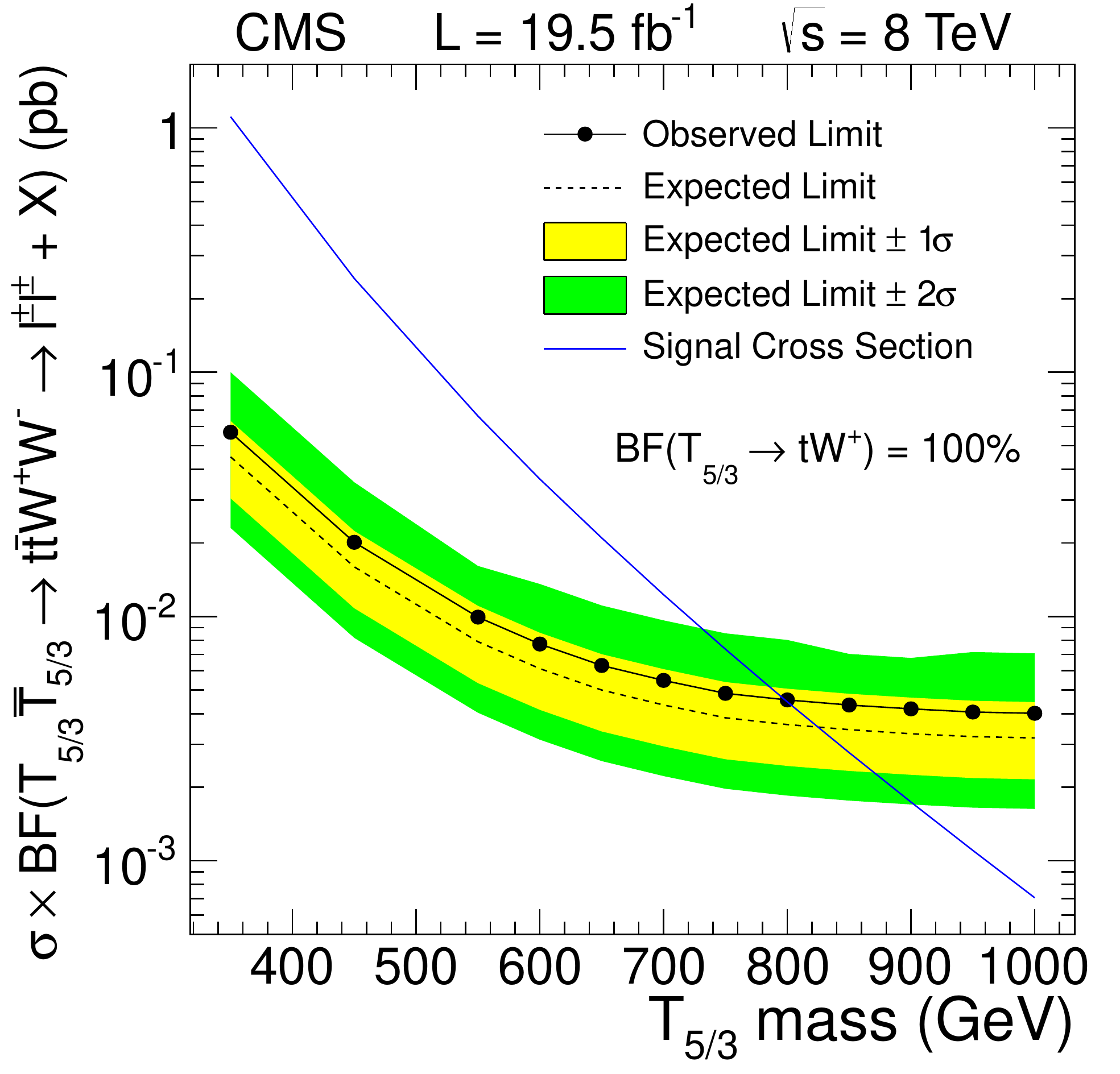}
\caption{Expected and observed 95\% CL limits on the $\Tft$ production cross section times the branching fraction for decay to same-sign dileptons.
         The $1\sigma$ and $2\sigma$ combined statistical and systematic expected variations are shown as yellow (light) and green (dark) bands, respectively.}
\label{fig:Limits}
\end{figure}

No significant excess is observed. Exclusion limits are computed at 95\% CL by using the \textsc{RooStats} implementation~\cite{RooStats} of the Bayesian approach. We use a cut-and-count method and compare the numbers of observed events with the numbers of expected signal and background events. A flat prior is used for the signal production cross section. The event yields from all lepton channels are combined when setting the limits. Upper bounds are set on the production cross section of heavy top-quark partners assuming a 100\% branching fraction (BF) for the decay $\Tft \to\cPqt\PW$. The resulting expected and observed limits are shown in Fig.~\ref{fig:Limits}. The expected  lower limit on the mass of the $\Tft$  is 830\GeV and the observed limit is 800\GeV.

The use of recently developed jet substructure techniques in this analysis for identifying boosted top quarks and \PW\ bosons enables us to  probe cross sections of $\Tft$ pair production that are between 10--20\% lower than would otherwise be possible for $\Tft$ masses in the range 800--1000\GeV. The reconstruction of the $\Tft$ mass
benefits as well and this can, in the event of a discovery in the future, be used to distinguish a $\Tft$ from other exotic particles which decay in a similar manner. (See \suppMaterial for figures illustrating the agreement of data with background predictions and showing the reconstructed top partner mass.)

In summary, a search for an exotic top partner with charge 5/3 in same-sign dileptonic events has been performed using $\usedLumi$ of data collected by the CMS experiment at $\sqrt{s}=8$\TeV.
No significant excess is observed in the data above the expected standard model background. An upper bound at the 95\% confidence level is set on the production cross section of heavy top-quark partners assuming a 100\% branching fraction for the decay $\Tft \to \cPqt\PW$, and masses below 800\GeV are excluded. This is the first limit on
$\Tft$ production from the LHC and it is  significantly more restrictive than the 365\GeV limit set by previous searches.

We congratulate our colleagues in the CERN accelerator departments for the excellent performance of the LHC and thank the technical and administrative staffs at CERN and at other CMS institutes for their contributions to the success of the CMS effort. In addition, we gratefully acknowledge the computing centres and personnel of the Worldwide LHC Computing Grid for delivering so effectively the computing infrastructure essential to our analyses. Finally, we acknowledge the enduring support for the construction and operation of the LHC and the CMS detector provided by the following funding agencies: BMWF and FWF (Austria); FNRS and FWO (Belgium); CNPq, CAPES, FAPERJ, and FAPESP (Brazil); MES (Bulgaria); CERN; CAS, MoST, and NSFC (China); COLCIENCIAS (Colombia); MSES (Croatia); RPF (Cyprus); MoER, SF0690030s09 and ERDF (Estonia); Academy of Finland, MEC, and HIP (Finland); CEA and CNRS/IN2P3 (France); BMBF, DFG, and HGF (Germany); GSRT (Greece); OTKA and NKTH (Hungary); DAE and DST (India); IPM (Iran); SFI (Ireland); INFN (Italy); NRF and WCU (Republic of Korea); LAS (Lithuania); CINVESTAV, CONACYT, SEP, and UASLP-FAI (Mexico); MSI (New Zealand); PAEC (Pakistan); MSHE and NSC (Poland); FCT (Portugal); JINR (Dubna); MON, RosAtom, RAS and RFBR (Russia); MESTD (Serbia); SEIDI and CPAN (Spain); Swiss Funding Agencies (Switzerland); NSC (Taipei); ThEPCenter, IPST, STAR and NSTDA (Thailand); TUBITAK and TAEK (Turkey); NASU (Ukraine); STFC (United Kingdom); DOE and NSF (USA).

\bibliography{auto_generated}   

\providecommand{\href}[2]{#2}\begingroup\raggedright\begin{thebibliography}{10}%
\makeatletter
\providecommand{\hrefCMSnoop }[0]{\@secondoftwo}%
\makeatother
\providecommand{\doi}{\texttt{doi:}\begingroup \urlstyle{tt}\Url}

\bibitem{DeSimone:2012fs}
\hrefCMSnoop {} {A.~De~Simone, O.~Matsedonskyi, R.~Rattazzi, and A.~Wulzer,
  ``{A first top partner hunter's guide}'',} \textit{ JHEP} \textbf{ 04} (2013)
  004,
  \href{http://dx.doi.org/10.1007/JHEP04(2013)004}{\doi{10.1007/JHEP04(2013)004}},
\href{http://www.arXiv.org/abs/1211.5663}{\texttt{ arXiv:1211.5663}}.

\bibitem{PhysRevD.81.075006}
\hrefCMSnoop {} {J.~Mrazek and A.~Wulzer, ``{A strong sector at the LHC: Top
  partners in same-sign dileptons}'',} \textit{ Phys. Rev. D} \textbf{ 81}
  (2010) 075006,
  \href{http://dx.doi.org/10.1103/PhysRevD.81.075006}{\doi{10.1103/PhysRevD.81.075006}},
  \href{http://www.arXiv.org/abs/0909.3977}{\texttt{ arXiv:0909.3977}}.

\bibitem{Dissertori:2010ug}
\hrefCMSnoop {} {G.~Dissertori, E.~Furlan, F.~Moortgat, and P.~Nef,
  ``{Discovery potential of top-partners in a realistic composite Higgs model
  with early LHC data}'',} \textit{ JHEP} \textbf{ 09} (2010) 019,
  \href{http://dx.doi.org/10.1007/JHEP09(2010)019}{\doi{10.1007/JHEP09(2010)019}},
\href{http://www.arXiv.org/abs/1005.4414}{\texttt{ arXiv:1005.4414}}.

\bibitem{Contino:2008hi}
\hrefCMSnoop {} {R.~Contino and G.~Servant, ``{Discovering the top partners at
  the LHC using same-sign dilepton final states}'',} \textit{ JHEP} \textbf{
  06} (2008) 026,
  \href{http://dx.doi.org/10.1088/1126-6708/2008/06/026}{\doi{10.1088/1126-6708/2008/06/026}},
\href{http://www.arXiv.org/abs/0801.1679}{\texttt{ arXiv:0801.1679}}.

\bibitem{Azatov:2011qy}
\hrefCMSnoop {} {A.~Azatov and J.~Galloway, ``{Light custodians and Higgs
  physics in composite models}'',} \textit{ Phys. Rev. D} \textbf{ 85} (2012)
  055013,
  \href{http://dx.doi.org/10.1103/PhysRevD.85.055013}{\doi{10.1103/PhysRevD.85.055013}},
\href{http://www.arXiv.org/abs/1110.5646}{\texttt{ arXiv:1110.5646}}.

\bibitem{Aad:2012tfa}
\hrefCMSnoop {} {{ ATLAS} Collaboration, ``{Observation of a new particle in
  the search for the Standard Model Higgs boson with the ATLAS detector at the
  LHC}'',} \textit{ Phys. Lett. B} \textbf{ 716} (2012) 1,
  \href{http://dx.doi.org/10.1016/j.physletb.2012.08.020}{\doi{10.1016/j.physletb.2012.08.020}},
\href{http://www.arXiv.org/abs/1207.7214}{\texttt{ arXiv:1207.7214}}.

\bibitem{higgspaper}
\hrefCMSnoop {} {{ CMS} Collaboration, ``{Observation of a new boson at a mass
  of 125 GeV with the CMS experiment at the LHC}'',} \textit{ Phys. Lett. B}
  \textbf{ 716} (2012) 30,
  \href{http://dx.doi.org/10.1016/j.physletb.2012.08.021}{\doi{10.1016/j.physletb.2012.08.021}},
\href{http://www.arXiv.org/abs/1207.7235}{\texttt{ arXiv:1207.7235}}.

\bibitem{Chatrchyan:2013lba}
\hrefCMSnoop {} {{ CMS} Collaboration, ``{Observation of a new boson with mass
  near 125 GeV in pp collisions at $\sqrt{s}$ = 7 and 8 TeV}'',} \textit{ JHEP}
  \textbf{ 06} (2013) 081,
  \href{http://dx.doi.org/10.1007/JHEP06(2013)081}{\doi{10.1007/JHEP06(2013)081}},
\href{http://www.arXiv.org/abs/1303.4571}{\texttt{ arXiv:1303.4571}}.

\bibitem{Aad:2013wqa}
\hrefCMSnoop {} {{ ATLAS} Collaboration, ``{Measurements of Higgs boson
  production and couplings in diboson final states with the ATLAS detector at
  the LHC}'',} \textit{ Phys. Lett. B} \textbf{ 726} (2013) 88,
  \href{http://dx.doi.org/10.1016/j.physletb.2013.08.010}{\doi{10.1016/j.physletb.2013.08.010}},
\href{http://www.arXiv.org/abs/1307.1427}{\texttt{ arXiv:1307.1427}}.

\bibitem{Aad:2013xqa}
\hrefCMSnoop {} {{ ATLAS} Collaboration, ``{Evidence for the spin-0 nature of
  the Higgs boson using ATLAS data}'',} \textit{ Phys. Lett. B} \textbf{ 726}
  (2013) 120,
  \href{http://dx.doi.org/10.1016/j.physletb.2013.08.026}{\doi{10.1016/j.physletb.2013.08.026}},
\href{http://www.arXiv.org/abs/1307.1432}{\texttt{ arXiv:1307.1432}}.

\bibitem{Chatrchyan:2012jja}
\hrefCMSnoop {} {{ CMS} Collaboration, ``{Study of the Mass and Spin-Parity of
  the Higgs Boson Candidate Via Its Decays to Z Boson Pairs}'',} \textit{ Phys.
  Rev. Lett.} \textbf{ 110} (2013) 081803,
  \href{http://dx.doi.org/10.1103/PhysRevLett.110.081803}{\doi{10.1103/PhysRevLett.110.081803}},
\href{http://www.arXiv.org/abs/1212.6639}{\texttt{ arXiv:1212.6639}}.

\bibitem{Chatrchyan:2013iaa}
\hrefCMSnoop {} {{ CMS} Collaboration, ``{Measurement of Higgs boson production
  and properties in the WW decay channel with leptonic final states}'',}
  \textit{ JHEP} \textbf{ 01} (2014) 096,
  \href{http://dx.doi.org/10.1007/JHEP01(2014)096}{\doi{10.1007/JHEP01(2014)096}},
\href{http://www.arXiv.org/abs/1312.1129}{\texttt{ arXiv:1312.1129}}.

\bibitem{Chatrchyan:2013mxa}
\hrefCMSnoop {} {{ CMS} Collaboration, ``{Measurement of the properties of a
  Higgs boson in the four-lepton final state}'',} (2014).
  \href{http://www.arXiv.org/abs/1312.5353}{\texttt{ arXiv:1312.5353}}.
Submitted to Phys. Rev. D.

\bibitem{Chatrchyan:2014vua}
\hrefCMSnoop {} {{ CMS} Collaboration, ``{Evidence for the direct decay of the
  125 GeV Higgs boson to fermions}'',} (2014).
  \href{http://www.arXiv.org/abs/1401.6527}{\texttt{ arXiv:1401.6527}}.
Submitted to Nature Physics.

\bibitem{Contino:2006qr}
\hrefCMSnoop {} {R.~Contino, L.~Da~Rold, and A.~Pomarol, ``{Light custodians in
  natural composite Higgs models}'',} \textit{ Phys. Rev. D} \textbf{ 75}
  (2007) 055014,
  \href{http://dx.doi.org/10.1103/PhysRevD.75.055014}{\doi{10.1103/PhysRevD.75.055014}},
\href{http://www.arXiv.org/abs/hep-ph/0612048}{\texttt{ arXiv:hep-ph/0612048}}.

\bibitem{Aaltonen:2009nr}
\hrefCMSnoop {} {{ CDF} Collaboration, ``{Search for New Bottomlike Quark Pair
  Decays $\mathrm{Q} \overline{\mathrm{Q}} \rightarrow (\mathrm{tW}^\mp)
  (\bar{\mathrm{t}}\mathrm{W}^\pm)$ in Same-Charge Dilepton Events}'',}
  \textit{ Phys. Rev. Lett.} \textbf{ 104} (2010) 091801,
  \href{http://dx.doi.org/10.1103/PhysRevLett.104.091801}{\doi{10.1103/PhysRevLett.104.091801}},
\href{http://www.arXiv.org/abs/0912.1057}{\texttt{ arXiv:0912.1057}}.

\bibitem{Aaltonen:2010js}
\hrefCMSnoop {} {{ CDF} Collaboration, ``{Exclusion of an Exotic Top Quark with
  $-4/3$ Electric Charge Using Soft Lepton Tagging}'',} \textit{ Phys. Rev.
  Lett.} \textbf{ 105} (2010) 101801,
  \href{http://dx.doi.org/10.1103/PhysRevLett.105.101801}{\doi{10.1103/PhysRevLett.105.101801}},
\href{http://www.arXiv.org/abs/1006.4597}{\texttt{ arXiv:1006.4597}}.

\bibitem{Chatrchyan:2008zzk}
\hrefCMSnoop {} {{ CMS} Collaboration, ``The {CMS} experiment at the {CERN}
  {LHC}'',} \textit{ JINST} \textbf{ 3} (2008) S08004,
  \href{http://dx.doi.org/10.1088/1748-0221/3/08/S08004}{\doi{10.1088/1748-0221/3/08/S08004}}.

\bibitem{Alwall:2011st}
J.~Alwall\hrefCMSnoop {} { {et~al.}, ``{MadGraph 5: going beyond.}'',} \textit{
  JHEP} \textbf{ 06} (2011) 128,
  \href{http://dx.doi.org/10.1007/JHEP06(2011)128}{\doi{10.1007/JHEP06(2011)128}},
\href{http://www.arXiv.org/abs/1106.0522}{\texttt{ arXiv:1106.0522}}.

\bibitem{Sjostrand:2006za}
\hrefCMSnoop {} {T.~Sj{\"o}strand, S.~Mrenna, and P.~Z. Skands, ``{PYTHIA} 6.4
  physics and manual'',} \textit{ JHEP} \textbf{ 05} (2006) 026,
  \href{http://dx.doi.org/10.1088/1126-6708/2006/05/026}{\doi{10.1088/1126-6708/2006/05/026}},
  \href{http://www.arXiv.org/abs/hep-ph/0603175}{\texttt{
  arXiv:hep-ph/0603175}}.

\bibitem{Pumplin:2002vw}
J.~Pumplin\hrefCMSnoop {} { {et~al.}, ``New generation of parton distributions
  with uncertainties from global {QCD} analysis'',} \textit{ JHEP} \textbf{ 07}
  (2002) 012,
  \href{http://dx.doi.org/10.1088/1126-6708/2002/07/012}{\doi{10.1088/1126-6708/2002/07/012}},
  \href{http://www.arXiv.org/abs/hep-ph/0201195v3}{\texttt{
  arXiv:hep-ph/0201195v3}}.

\bibitem{UEPAS}
\hrefCMSnoop {} {{ CMS} Collaboration, ``Measurement of the underlying event
  activity at the LHC with $\sqrt{s\ }=7$ TeV and Comparison with $\sqrt{s} =
  0.9$ TeV'',} \textit{ JHEP} \textbf{ 09} (2011) 109,
  \href{http://dx.doi.org/10.1007/JHEP09(2011)109}{\doi{10.1007/JHEP09(2011)109}},
  \href{http://www.arXiv.org/abs/1107.0330v1}{\texttt{ arXiv:1107.0330v1}}.

\bibitem{GEANT}
\hrefCMSnoop {} {{ GEANT4} Collaboration, ``{GEANT4}---a simulation toolkit'',}
  \textit{ Nucl. Instrum. Meth. A} \textbf{ 506} (2003) 250,
\href{http://dx.doi.org/10.1016/S0168-9002(03)01368-8}{\doi{10.1016/S0168-9002(03)01368-8}}.

\bibitem{Czakon:2013goa}
\hrefCMSnoop {} {M.~Czakon, P.~Fiedler, and A.~Mitov, ``{The total top quark
  pair production cross-section at hadron colliders through
  $\mathcal{O}(\alpha_\mathrm{S}^{4})$}'',} \textit{ Phys. Rev. Lett.} \textbf{
  110} (2013) 252004,
  \href{http://dx.doi.org/10.1103/PhysRevLett.110.252004}{\doi{10.1103/PhysRevLett.110.252004}},
\href{http://www.arXiv.org/abs/1303.6254}{\texttt{ arXiv:1303.6254}}.

\bibitem{Garzelli:2012bn}
\hrefCMSnoop {} {M.~V. Garzelli, A.~Kardos, C.~G. Papadopoulos, and
  Z.~Trocsanyi, ``{$\ttbar\PW^\pm$ and $\ttbar\cPZ$ hadroproduction at NLO
  accuracy in QCD with parton shower and hadronization effects}'',} \textit{
  JHEP} \textbf{ 11} (2012) 056,
  \href{http://dx.doi.org/10.1007/JHEP11(2012)056}{\doi{10.1007/JHEP11(2012)056}},
\href{http://www.arXiv.org/abs/1208.2665}{\texttt{ arXiv:1208.2665}}.

\bibitem{MCFM}
\hrefCMSnoop {} {J.~M. Campbell, R.~K. Ellis, and C.~Williams, ``Vector boson
  pair production at the {LHC}'',} \textit{ JHEP} \textbf{ 07} (2011) 018,
  \href{http://dx.doi.org/10.1007/JHEP07(2011)018}{\doi{10.1007/JHEP07(2011)018}},
  \href{http://www.arXiv.org/abs/1105.0020}{\texttt{ arXiv:1105.0020}}.

\bibitem{pf}
\href {http://cdsweb.cern.ch/record/1279347} {{CMS Collaboration},
  ``Commissioning of the particle-flow event reconstruction with leptons from
  {J}/$\psi$ and {W} decays at 7 {TeV}'',} CMS Physics Analysis Summary
  {CMS-PAS-PFT-10-003}, 2010.

\bibitem{pf2}
\href {http://cdsweb.cern.ch/record/1194487} {{CMS Collaboration},
  ``Particle-Flow Event Reconstruction in CMS and Performance for Jets, Taus,
  and Missing ET'',} CMS Physics Analysis Summary {CMS-PAS-PFT-09-001}, 2009.

\bibitem{ctf}
\href {https://cdsweb.cern.ch/record/1258204} {{CMS Collaboration}, ``Tracking
  and Vertexing Results from First Collisions'',} CMS Physics Analysis Summary
  {CMS-PAS-TRK-10-001}, 2010.

\bibitem{gsf}
\hrefCMSnoop {} {W.~Adam, R.~Fr{\"u}hwirth, A.~Strandlie, and T.~Todorov,
  ``{Reconstruction of electrons with the Gaussian sum filter in the CMS
  tracker at LHC}'',} \textit{ J. Phys. G} \textbf{ 31} (2005) N9,
  \href{http://dx.doi.org/10.1088/0954-3899/31/9/N01}{\doi{10.1088/0954-3899/31/9/N01}},
\href{http://www.arXiv.org/abs/physics/0306087}{\texttt{
  arXiv:physics/0306087}}.

\bibitem{CMS:2011aa}
\hrefCMSnoop {} {{ CMS} Collaboration, ``{Measurement of the inclusive $\rm{W}$
  and $\rm{Z}$ production cross sections in pp collisions at $\sqrt{s}=7$
  TeV}'',} \textit{ JHEP} \textbf{ 10} (2011) 132,
  \href{http://dx.doi.org/10.1007/JHEP10(2011)132}{\doi{10.1007/JHEP10(2011)132}},
\href{http://www.arXiv.org/abs/1107.4789}{\texttt{ arXiv:1107.4789}}.

\bibitem{CMS-PAS-JME-09-001}
\href {http://cds.cern.ch/record/1194489} {{ CMS} Collaboration, ``A
  Cambridge-Aachen (C-A) based jet algorithm for boosted top-jet tagging'',}
  CMS Physics Analysis Summary {CMS-PAS-JME-09-001}, 2009.

\bibitem{Wobisch:1998wt}
\href {http://www.desy.de/~heramc/proceedings/wg40/} {M.~Wobisch and
  T.~Wengler, ``{Hadronization corrections to jet cross-sections in deep
  inelastic scattering}'',} (1999).
  \href{http://www.arXiv.org/abs/hep-ph/9907280}{\texttt{
  arXiv:hep-ph/9907280}}.
DESY-PROC 1999-02.

\bibitem{Dokshitzer:1997in}
\hrefCMSnoop {} {Y.~L. Dokshitzer, G.~D. Leder, S.~Moretti, and B.~R. Webber,
  ``{Better jet clustering algorithms}'',} \textit{ JHEP} \textbf{ 08} (1997)
  001,
  \href{http://dx.doi.org/10.1088/1126-6708/1997/08/001}{\doi{10.1088/1126-6708/1997/08/001}},
\href{http://www.arXiv.org/abs/hep-ph/9707323}{\texttt{ arXiv:hep-ph/9707323}}.

\bibitem{Cacciari:2011ma}
\hrefCMSnoop {} {M.~Cacciari, G.~P. Salam, and G.~Soyez, ``{FastJet user
  manual}'',} \textit{ Eur. Phys. J. C} \textbf{ 72} (2012) 1896,
  \href{http://dx.doi.org/10.1140/epjc/s10052-012-1896-2}{\doi{10.1140/epjc/s10052-012-1896-2}},
\href{http://www.arXiv.org/abs/1111.6097}{\texttt{ arXiv:1111.6097}}.

\bibitem{CMS-PAS-JME-10-013}
\href {http://cds.cern.ch/record/1333700} {{ CMS} Collaboration, ``Study of Jet
  Substructure in pp Collisions at 7 {TeV} in {CMS}'',} CMS Physics Analysis
  Summary {CMS-PAS-JME-10-013}, 2010.

\bibitem{Ellis:2009su}
\hrefCMSnoop {} {S.~D. Ellis, C.~K. Vermilion, and J.~R. Walsh, ``{Techniques
  for improved heavy particle searches with jet substructure}'',} \textit{
  Phys. Rev. D} \textbf{ 80} (2009) 051501,
  \href{http://dx.doi.org/10.1103/PhysRevD.80.051501}{\doi{10.1103/PhysRevD.80.051501}},
\href{http://www.arXiv.org/abs/0903.5081}{\texttt{ arXiv:0903.5081}}.

\bibitem{Ellis:2009me}
\hrefCMSnoop {} {S.~D. Ellis, C.~K. Vermilion, and J.~R. Walsh,
  ``{Recombination algorithms and jet substructure: Pruning as a tool for heavy
  particle searches}'',} \textit{ Phys. Rev. D} \textbf{ 81} (2010) 094023,
  \href{http://dx.doi.org/10.1103/PhysRevD.81.094023}{\doi{10.1103/PhysRevD.81.094023}},
\href{http://www.arXiv.org/abs/0912.0033}{\texttt{ arXiv:0912.0033}}.

\bibitem{CMS-PAS-JME-13-006}
\href {http://cds.cern.ch/record/1577417} {{ CMS} Collaboration, ``Identifying
  Hadronically Decaying Vector Bosons Merged into a Single Jet'',} CMS Physics
  Analysis Summary {CMS-PAS-JME-13-006}, 2013.

\bibitem{Cacciari:2008gp}
\hrefCMSnoop {} {M.~Cacciari, G.~P. Salam, and G.~Soyez, ``{The anti-$k_t$ jet
  clustering algorithm}'',} \textit{ JHEP} \textbf{ 04} (2008) 063,
  \href{http://dx.doi.org/10.1088/1126-6708/2008/04/063}{\doi{10.1088/1126-6708/2008/04/063}},
\href{http://www.arXiv.org/abs/0802.1189}{\texttt{ arXiv:0802.1189}}.

\bibitem{JINSTJetPaper}
\hrefCMSnoop {} {{ CMS} Collaboration, ``{Determination of jet energy
  calibration and transverse momentum resolution in CMS}'',} \textit{ JINST}
  \textbf{ 6} (2011) 11002,
  \href{http://dx.doi.org/10.1088/1748-0221/6/11/P11002}{\doi{10.1088/1748-0221/6/11/P11002}},
  \href{http://www.arXiv.org/abs/1107.4277}{\texttt{ arXiv:1107.4277}}.

\bibitem{Fastjet1}
\hrefCMSnoop {} {M.~Cacciari and G.~P. Salam, ``{Pileup subtraction using jet
  areas}'',} \textit{ Phys. Lett. B} \textbf{ 659} (2008) 119,
\href{http://dx.doi.org/10.1016/j.physletb.2007.09.077}{\doi{10.1016/j.physletb.2007.09.077}}.

\bibitem{Fastjet2}
\hrefCMSnoop {} {M.~Cacciari, G.~P. Salam, and G.~Soyez, ``The catchment area
  of jets'',} \textit{ JHEP} \textbf{ 04} (2008) 005,
\href{http://dx.doi.org/10.1088/1126-6708/2008/04/005}{\doi{10.1088/1126-6708/2008/04/005}}.

\bibitem{Chatrchyan:2013lca}
\hrefCMSnoop {} {{ CMS} Collaboration, ``{Searches for new physics using the
  $\ttbar$ invariant mass distribution in $pp$ collisions at $\sqrt{s}=8$
  TeV}'',} \textit{ Phys. Rev. Lett.} \textbf{ 111} (2013) 211804,
  \href{http://dx.doi.org/10.1103/PhysRevLett.111.211804}{\doi{10.1103/PhysRevLett.111.211804}},
\href{http://www.arXiv.org/abs/1309.2030}{\texttt{ arXiv:1309.2030}}.

\bibitem{susy2011}
\hrefCMSnoop {} {{ CMS} Collaboration, ``{Search for new physics in events with
  same-sign dileptons and b-tagged jets in pp collisions at $\sqrt{s}=7$
  TeV}'',} \textit{ JHEP} \textbf{ 08} (2012) 110,
  \href{http://dx.doi.org/10.1007/JHEP08(2012)110}{\doi{10.1007/JHEP08(2012)110}},
\href{http://www.arXiv.org/abs/1205.3933}{\texttt{ arXiv:1205.3933}}.

\bibitem{Chatrchyan:2011wba}
\hrefCMSnoop {} {{ CMS} Collaboration, ``{Search for new physics with same-sign
  isolated dilepton events with jets and missing transverse energy at the
  LHC}'',} \textit{ JHEP} \textbf{ 06} (2011) 077,
  \href{http://dx.doi.org/10.1007/JHEP06(2011)077}{\doi{10.1007/JHEP06(2011)077}},
\href{http://www.arXiv.org/abs/1104.3168}{\texttt{ arXiv:1104.3168}}.

\bibitem{WZAtlas2012}
\hrefCMSnoop {} {{ ATLAS} Collaboration, ``Measurement of the {WZ} production
  cross section and limits on \ anomalous triple gauge couplings in
  proton-proton collisions at $\sqrt{s}$ = 7 TeV with the {ATLAS} detector'',}
  \textit{ Phys. Lett. B} \textbf{ 709} (2012) 341,
  \href{http://dx.doi.org/10.1016/j.physletb.2012.02.053}{\doi{10.1016/j.physletb.2012.02.053}},
  \href{http://www.arXiv.org/abs/1111.5570}{\texttt{ arXiv:1111.5570}}.

\bibitem{Campbell:2012dh}
\hrefCMSnoop {} {J.~M. Campbell and R.~K. Ellis, ``{$\mathrm{t}
  \bar{\mathrm{t}} \mathrm{W^{\pm}}$ production and decay at NLO}'',} \textit{
  JHEP} \textbf{ 07} (2012) 052,
  \href{http://dx.doi.org/10.1007/JHEP07(2012)052}{\doi{10.1007/JHEP07(2012)052}},
\href{http://www.arXiv.org/abs/1204.5678}{\texttt{ arXiv:1204.5678}}.

\bibitem{Chatrchyan:2012paa}
\hrefCMSnoop {} {{ CMS} Collaboration, ``{Search for new physics in events with
  same-sign dileptons and b jets in pp collisions at $\sqrt{s}=8$ TeV}'',}
  \textit{ JHEP} \textbf{ 03} (2013) 037,
  \href{http://dx.doi.org/10.1007/JHEP03(2013)037}{\doi{10.1007/JHEP03(2013)037}},
  \href{http://www.arXiv.org/abs/1212.6194}{\texttt{ arXiv:1212.6194}}.
Erratum at \doi{10.1007/JHEP07(2013)041}.

\bibitem{CMS-PAS-LUM-13-001}
\href {http://cds.cern.ch/record/1482193} {{ CMS} Collaboration, ``CMS
  luminosity based on pixel cluster counting - Summer 2013 Update'',} CMS
  Physics Analysis Summary {CMS-PAS-LUM-13-001}, 2013.

\bibitem{RooStats}
L.~Moneta\href
  {http://pos.sissa.it/archive/conferences/093/057/ACAT2010_057.pdf} {
  {et~al.}, ``The {R}oo{S}tats {P}roject'',} in \textit{ The 13$^\text{th}$
  International Workshop on Advanced Computing and Analysis Techniques in
  Physics Research (ACAT2010)}.
\newblock SISSA, 2010.
\newblock \href{http://www.arXiv.org/abs/1009.1003}{\texttt{ arXiv:1009.1003}}.

\end{thebibliography}\endgroup

\ifthenelse{\boolean{cms@external}}{}{
\clearpage
\appendix
\section{Additional Kinematic Distributions\label{app:suppMat}}
\providecommand{\Tft}{\ensuremath{\cmsSymbolFace{T}_{5/3}}\xspace}
Additional figures are provided to illustrate the agreement of data with background predictions and to show the reconstruction
of the $\Tft$ mass. Figure~\ref{fig:preSelHT} gives the \HT distribution after the requirement of same-sign dileptons, the
\cPZ-boson veto, and a requirement of at least two jets. Figure~\ref{fig:T53Mass} shows the distribution of the reconstructed
$\Tft$ mass.
The mass of the $\Tft$ quark can be reconstructed when the decay of one of the top-quark partners leads to same-sign
leptons and the decay of the second partner is fully hadronic. The selection for this distribution requires same-sign leptons,
the \cPZ-boson veto, and the presence of enough jets to reconstruct a top quark and a $\PW$ boson using any combination of
top-quark jets, \PW-boson jets, and AK5 jets. Top-quark and \PW-boson jets are assumed to correspond to the respective particles,
while combinations of AK5 jets are required to have an invariant mass consistent with the particles from which they originate.
\begin{figure}[htb]
\centering
\includegraphics[width=0.48\textwidth]{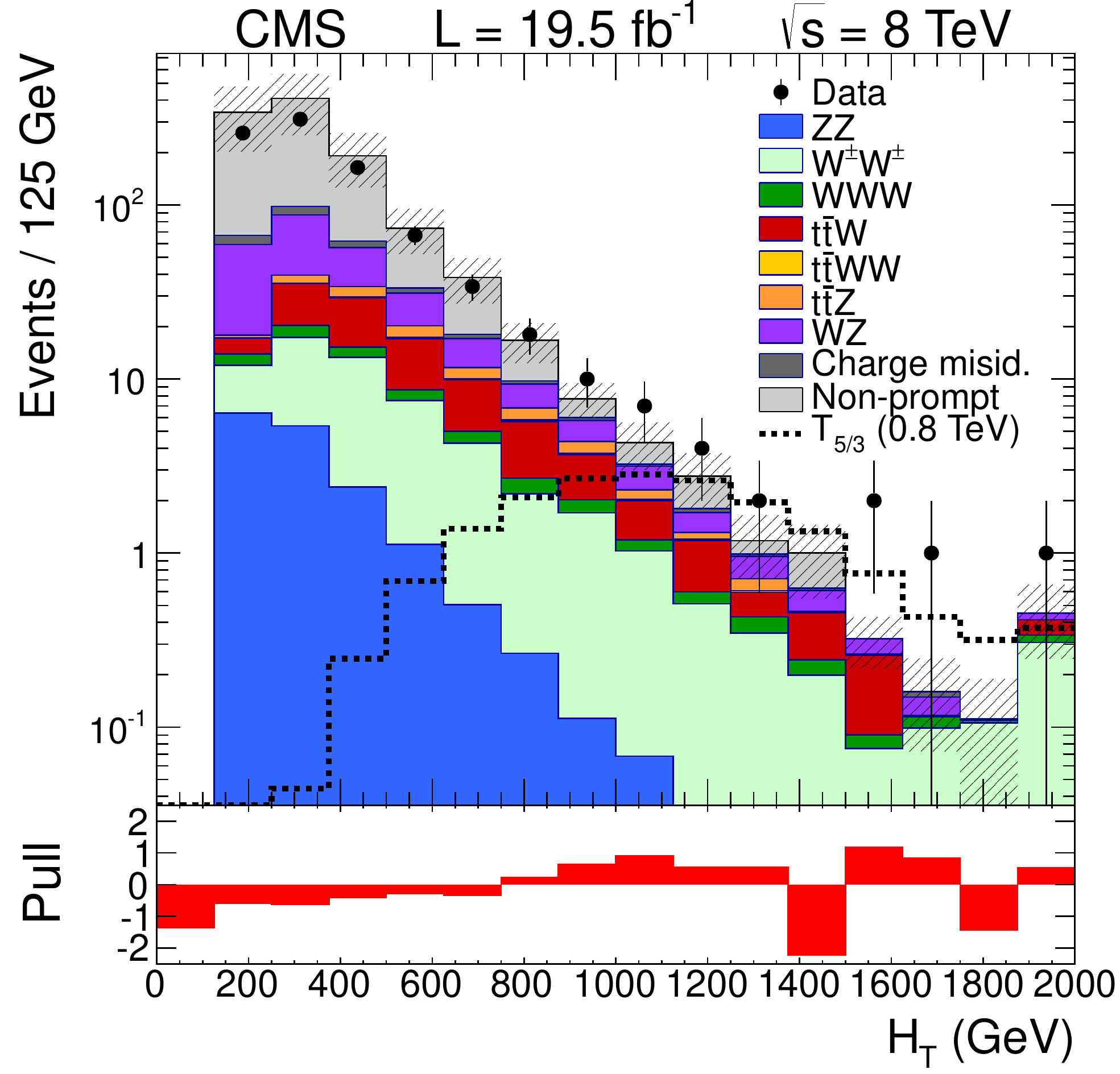}
\caption{The distribution of \HT for all channels combined, after after the requirement of same-sign dileptons, the \cPZ-boson veto,
and a requirement of at least two jets. The shaded band represents the total uncertainty in the predicted backgrounds.
The final bin includes all overflow events. The pull is defined as the difference between the observed and expected values divided by
the total uncertainty.}
\label{fig:preSelHT}
\end{figure}
\begin{figure}[htb]
\centering
\includegraphics[width=0.48\textwidth]{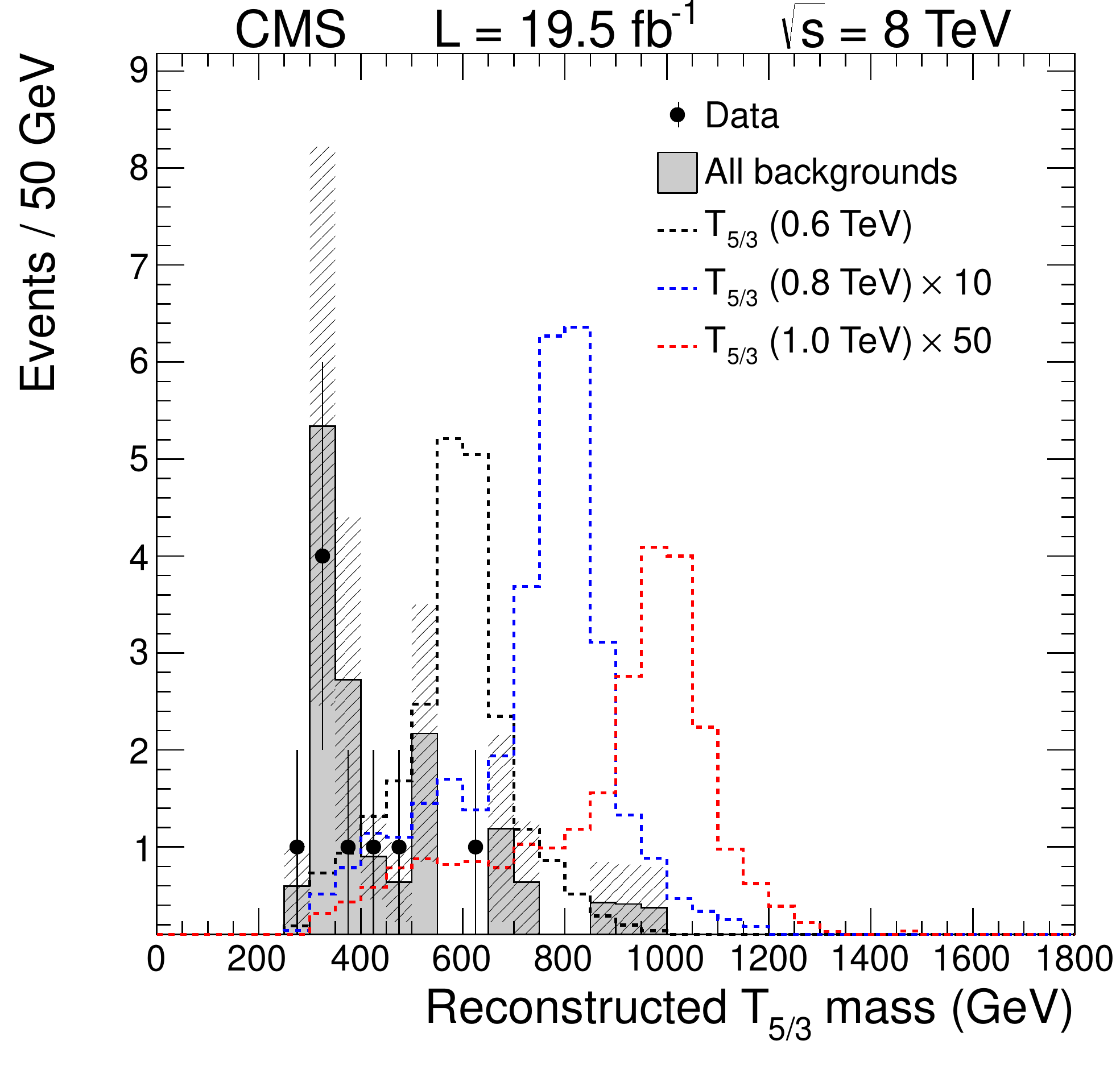}
\caption{The distribution of the reconstructed $\Tft$ mass for the data, the background, and three signal mass points. The background
is dominated by the contribution due to non-prompt leptons and also contains much smaller contributions due to $\ttbar \PW$,
$\ttbar\PW\PW$, $\ttbar\cPZ$, $\PW\PW\PW$, same-sign $\PW\PW$, and charge misidentification.
The shaded band represents the total uncertainty in the predicted background.}
\label{fig:T53Mass}
\end{figure}

This selection differs from the one used for computing the 95\% CL limit: not every event used for the limit setting has jet combinations
that result in a top quark and a \PW~boson and there is no \HT requirement in the selection used for the $\Tft$ mass reconstruction.
The $\Tft$ mass distribution is consistent with the exclusion of the $\Tft$ in this mass range: no excess is observed and the jet 
content of the data is different from that of the signal. The nine data events contain only AK5 jets whereas even at the relatively
low mass of 600 \GeV, 75\% of $\Tft$ events for which the mass can be reconstructed are expected to have at least one \PW-boson jet
and 17\% are expected to have at least one top-quark jet.

}

\cleardoublepage \section{The CMS Collaboration \label{app:collab}}\begin{sloppypar}\hyphenpenalty=5000\widowpenalty=500\clubpenalty=5000\textbf{Yerevan Physics Institute,  Yerevan,  Armenia}\\*[0pt]
S.~Chatrchyan, V.~Khachatryan, A.M.~Sirunyan, A.~Tumasyan
\vskip\cmsinstskip
\textbf{Institut f\"{u}r Hochenergiephysik der OeAW,  Wien,  Austria}\\*[0pt]
W.~Adam, T.~Bergauer, M.~Dragicevic, J.~Er\"{o}, C.~Fabjan\cmsAuthorMark{1}, M.~Friedl, R.~Fr\"{u}hwirth\cmsAuthorMark{1}, V.M.~Ghete, C.~Hartl, N.~H\"{o}rmann, J.~Hrubec, M.~Jeitler\cmsAuthorMark{1}, W.~Kiesenhofer, V.~Kn\"{u}nz, M.~Krammer\cmsAuthorMark{1}, I.~Kr\"{a}tschmer, D.~Liko, I.~Mikulec, D.~Rabady\cmsAuthorMark{2}, B.~Rahbaran, H.~Rohringer, R.~Sch\"{o}fbeck, J.~Strauss, A.~Taurok, W.~Treberer-Treberspurg, W.~Waltenberger, C.-E.~Wulz\cmsAuthorMark{1}
% ...
\vskip\cmsinstskip
% ...
1:~~Also at Vienna University of Technology, Vienna, Austria\\
2:~~Also at CERN, European Organization for Nuclear Research, Geneva, Switzerland\\
% ...

\end{sloppypar}
\end{document}